\documentclass[aps,a4paper,10pt,twocolumn,nofootinbib]{revtex4}

\usepackage{amsmath}
\usepackage{amsthm}
\usepackage{amssymb}
\usepackage{tikz}
\usepackage[mathscr]{euscript}
\usepackage{hyperref}

\usepackage{tikz}
\usepackage{flowchart}
\usetikzlibrary{arrows}
\usetikzlibrary{external}

\newcommand{\XWEB}[1]{\href{#1}{#1}}
\newcommand{\XDOI}[1]{\href{http://dx.doi.org/#1}{doi:#1}}
\newcommand{\XARXIV}[1]{\href{http://arxiv.org/abs/#1}{arXiv:#1}}

\def\Permittivity{\varepsilon}
\def\cspeed{c}

\def\Radius{r}
\def\Radiusz{R(z)}
\def\Length{L}

\def\Cutofff{\lambda}
\def\CutofffSim{\Cutofff_{\textup{s}}}
\def\CutofffSimr{\CutofffSim(\Radius)}

\def\Cutoff{\Cutofff^2}
\def\Cutoffr{\Cutoff(\Radius)}
\def\CutoffSim{\Cutoff_{\textup{s}}}
\def\CutoffSimr{\CutoffSim(\Radius)}
\def\Cutoffz{\Lambda^2(z)}

\def\Bpolarv{\beta}
\def\pfa{4\pi^2}

\def\wavefq{f}       
\def\waveka{\kappa}  
\def\PlasmaFq{\wavefq_\textup{p}}

\newcommand{\Units}[1]{\textup{#1}}

\definecolor{XRED}{rgb}{0.70, 0.01, 0.01}
\definecolor{XBLUE}{rgb}{0.01, 0.01, 0.70}

\def\UPDATED#1{#1}
\def\UPDATES#1{#1}

\begin{document}

\title{Customized longitudinal electric field profiles:\\
 using spatial dispersion in dielectric wire arrays}

\author{Taylor Boyd$^1$}

\author{Jonathan Gratus$^{2,1}$}
\homepage[]{https://orcid.org/0000-0003-1597-6084}
\email[\hphantom{.}~]{j.gratus@lancaster.ac.uk}

\author{Paul Kinsler$^{2,1}$}
\homepage[]{https://orcid.org/0000-0001-5744-8146}

\author{Rosa Letizia$^3$}
\homepage[]{https://orcid.org/0000-0002-1664-2265}

\affiliation{
    $^1$Cockcroft Institute, Sci-Tech Daresbury, 
      Daresbury WA4 4AD, United Kingdom}
\affiliation{
    $^2$Physics Department, Lancaster University, 
      Lancaster LA1 4YB, United Kingdom}
\affiliation{
    $^3$Engineering Department, Lancaster University, 
      Lancaster LA1 4YW, United Kingdom}

\begin{abstract}
We show how spatial dispersion can be used as a mechanism 
 to customize the longitudinal profiles 
 of electric fields inside modulated wire media, 
 using a fast and remarkably accurate 1D inhomogeneous model.
This customization gives fine control of the sub-wavelength behaviour
 of the field, 
 as has been achieved recently for transverse fields
 in simpler slotted-slab media.
Our scheme avoids any necessity to run a long series
 of computationally intensive 3D simulations of specific structures, 
 in order to iteratively converge (or brute-force search)
 to an empirical `best-performance' design according to an abstract
 figure-of-merit.
Instead, 
 if supplied with an `ideal waveform' profile,
 we could now calculate how to construct it directly.
Notably, 
 and unlike most work on photonic crystal structures,
 our focus is specifically on the field profiles
 because of their potential utility,
 rather than other issues such as band-gap control, 
 and/or transmission and reflection coefficients.
%
%
%
%
\end{abstract}

\date{\today}


\maketitle

\section{Introduction}
\label{Intro}

Detailed control over electromagnetic field profiles
 (the field strength variation through space)
 and other electromagnetic properties is an invaluable ability. 
In existing accelerator technology,
 crab cavities \cite{VerduAndres-BBCWX-2016nppp}
 are used to provide a tailored field profile
 for control of the particle bunch;
 and photonic crystals are used for band gap,
 transmission,
 and dispersion control \cite{Russell-B-1999jlt,Russell-BL-1995natoasi}.
More recently,
 the emerging multidisciplinary field of metaphotonics \cite{Baev-PASW-2015pr}
 attempts to further such aims with subwavelength (meta)material design,
 using both electric and magnetic interactions and their cross-coupling.
Such technology promises a remarkable degree of control,
 enabling zero (or negative) electromagnetic response,
 chirality control,
 and cloaking.

Our interest here is to be able to design any electric field profile
 on a wavelength scale; 
 that is,
 to sculpt the carrier wave of a propagating field itself. 
This idea has previously been investigated
 in the context of
 nonlinearity-induced carrier shocking \cite{Radnor-CKN-2008pra}
 and by harmonic synthesis \cite{Chan-HLKLLPP-2011s,Cox-PSLK-2012ol}.
More
 recent work has outlined the potential for mode profiling
 using modulated dielectric slabs
 \cite{Gratus-KLB-2017apa-malaga,Gratus-PLB-2017jpc}, 
 but only for transverse field modes.
However, 
 here we aim to control the \emph{longitudinal} field component, 
 not the transverse one.
For this we use dielectric wire medium, 
 a choice crucial to our success, 
 because it is spatially dispersive 
 and 
 natively supports modes in which the longitudinal electric field 
 component can dominate the transverse ones.

The method outlined in the following sections
 allows for the rapid and accurate generation of any chosen field profile
 in the wire media structure; 
 further, 
 once some uniform-radius reference simulations are done in 3D, 
 all further design need only be based on a simple 1D model.
This high level of control could be used to
 tailor field profiles for specific applications
 such as enhancing ionization
 in high harmonic generation \cite{Radnor-CKN-2008pra}.
One could also imagine the creation of electric fields with high gradients
 without the usually associated high peak field values (i.e a flattened mode),
 which can lead to unwanted nonlinear effects. 
Alternatively, 
 an electromagnetic profile with a more pronounced peak
 would be useful in signal processing
 as it would make for easier detection amongst any background noise.

However, 
 one significant area of our interest is that of accelerator applications
 such as the shaping of electron bunches
 or uses in the plasma ionisation
 involved in laser wakefield accelerators
 \cite{Piot-SPR-2011prstab,Albert-TMBCFLNVN-2014ppcf}.
This is because of our new ability to gain detailed control
 over the field profiles and gradients,
 and not just aim at a general redistribution of field intensity 
 as a means to moderate losses or heating.
For example, 
 in accelerator applications,
 an RF pillbox cavity is used as a way
 to increase the available accelerating voltage with a minimum power loss.
Typical structural features involve the use of a ``nose-cone''
 profile to concentrate the electric field
 near the particle beam passage; 
 but the design
 involves
 an iterative process of numerically optimizing 
 the cavity figures of merit
 based on full 3D simulations \cite{LINAC-Larranaga-2010,LINAC-Shin-2014}.
This is fundamentally different from our method, 
 which requires only a 1D specification of the field profile, 
 and then --
 without iteration --
 gives the radius variation needed.

Of course, 
 to be truly useful in accelerator applications, 
 we would prefer the field profiles to be co-moving
 with the particle bunches, 
 or to have control over the time domain profiles.
In what we present here the profiles have stationary spatial profiles
 and ordinary sinusoidal behavior in time; 
 nevertheless our results show the first important step towards
 our eventual goal.

\begin{figure}[h]
{\centering
\resizebox{0.99\columnwidth}{!}{
\includegraphics[height=0.13\columnwidth]{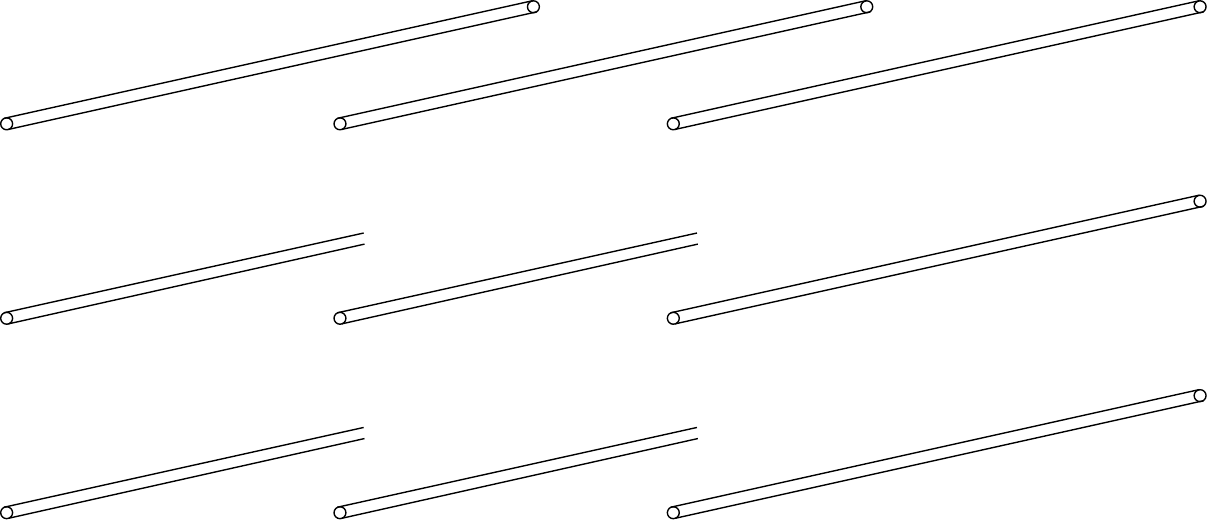}
\quad
\includegraphics[height=0.13\columnwidth]{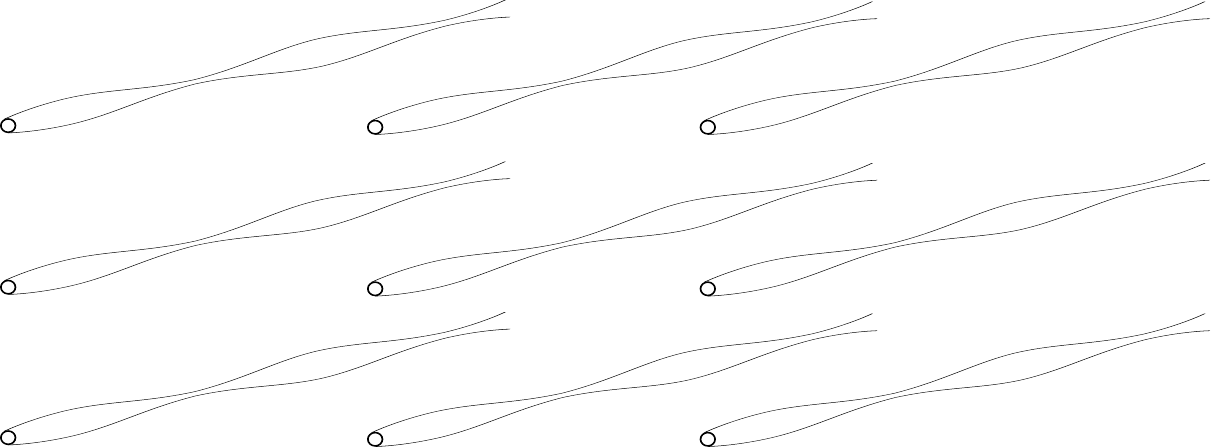}}}
\caption{A diagram showing 
 a standard uniform-radius wire medium {(left)}
 compared with a varying-radius wire medium {(right)}
 for wave profile shaping.
By understanding and characterizing the properties of a uniform medium, 
 we can use them as a basis to predict the behaviour 
 of varying-radius medium, 
 and so perform sub-wavelength mode profile shaping.
The important feature of such wire media is that they naturally support
 modes with longitudinal electric fields, 
 i.e. fields parallel to the wires themselves.
}
\label{fig-Wires}
\end{figure}

\begin{figure}
\centering
\includegraphics[width=0.65\columnwidth]{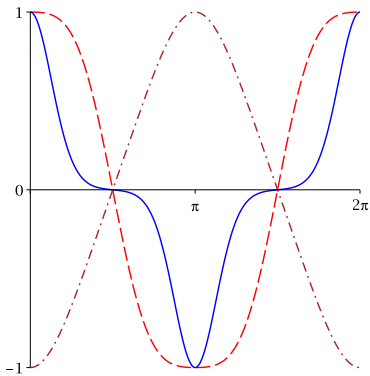}
\caption{Some potentially interesting solutions to Mathieu's equations:
 a wave with 
 a flat-topped profile
 (dashed line, $q=0.8$ and $a = 1.7118$);
 a triangular profile
 (dot-dashed line, $q=-0.329$ and $a = 0.6580$);
 and one with pronounced peaks
 (solid line, with parameters $q=-10.0$ and $a = -13.9366$).
}
\label{fig-Mathieu}
\end{figure}

\UPDATED{We will show here that field modes 
 with a \emph{longitudinal} electric field 
 exist 
 in dielectric wire media.
Starting with results based on uniform-radius wires, 
 we show that by implementing a design where
 it includes a variation of the radius
 (see fig. \ref{fig-Wires}),
 we can generate a modulation of the mode profile.}
In particular, 
 we focus on periodic variations of the wire radius,
 where the electric field can then be shown to obey 
 Mathieu's equation \cite{WfmMathWorld-Mathieu}, 
 i.e.
~
\begin{equation}
\frac{d^{2}y}{d\sigma^{2}}+(a-2q\cos(2\sigma))y=0
.
\label{eqn-Mathieu}
\end{equation}
This is a well established differential equation
 with a set of known solutions, 
 where the free parameters $q$ and $a$ control the properties of the solution;
 \UPDATED{sample solutions
 which we will use as target mode profiles 
 are shown on fig. \ref{fig-Mathieu}.}
In particular,
 for a given $q$ there always exists 
 a discrete spectrum of $a$ values
\UPDATED{ such that the solutions $y(\sigma)$ are periodic
 over $2\pi$ \cite{WfmMathWorld-Mathieu}.
In the trivial case where $q=0$, 
 we find that $a \in \{0,1,4,9,...\}$.}
In this work the focus will be purely on periodic solutions
 of Mathieu's equation, 
 although we expect that our method can be applied to treat
 non-periodic solutions,
 and could even be generalized beyond the Mathieu equation to 
 generate a range of non-periodic field profiles.

As can be seen in fig. \ref{fig-Mathieu}
 some of the periodic solutions of Mathieu's equation
 fit the description of the potentially desirable profiles discussed earlier. 
The flatter profile which could minimise the harmful impact
 of nonlinear effects,
 whereas the profile with more pronounced peaks
 could allow easier detection above backgound noise.
Either might be useful to control the field experienced 
 by particles located within a larger bunch in an accelerator, 
 ensuring either a more uniform acceleration or an enhanced chirp.
Our challenge here is to show how modulations in the wire radii 
 can be mapped onto a sufficiently simple electric field equation,
 which then also mimics Mathieu's equation; 
 \UPDATED{this is therefore similar in principle 
 to the rather more straightforward mapping from
 the physics to Mathieu's equation
 possible in the context
 of Josephson junctions \cite{Wilkinson-VGC-2017arxiv}.}
This is achieved in section \ref{Profiling}, 
 which shows how longitudinal modes obeying Mathieu's equation are supported, 
 thus allowing use to generate Mathieu-like mode profiles
 in 3D structures
 using an efficient 1D design process.
In addition to our theoretical modelling, 
 we have validated our procedure for a full 3D wire medium
 by means of simulations using CST Microwave Studio \cite{citeCST}.

\section{Inhomogeneous and spatially dispersive 1D model}\label{S-1dmodel}

Wire media are a well known class of metamaterials
 consisting of a lattice of parallel wires,
 with the radius $r$ of the wires
 being small compared to their spacing \cite{Gratus-M-2015jo}.
It should be noted that wire media act as metamaterials
 when the Bragg scattering wavelength
 is less than the Mie scattering wavelength
 (i.e when low permittivity rods are tightly spaced
 or rods with large spacing have high permittivity). 
As the ratio between permittivity and lattice spacing decreases
 wire media start to enter the regime of photonic crystals \cite{Rybin-FSBKL-2015nc}. 
One of the unique features of these structures
 is that they support purely electric longitudinal modes
 which have a plasma-like dispersion relation.

Wire media are complex three-dimensional inhomogeneous media,
  which makes them difficult to study
  without introducing a simpler model for the system. 
The model we will use approximates the lattice of wires
  as a one-dimensional inhomogeneous media which exhibits spatial dispersion. 
Using this model will greatly simplify our efforts
  to study the behaviour and manipulation of the profile
  of longitudinal electric modes in wire media. 
If we can then make predictions about these modes
 which are validated in full 3D simulations,
 it will also validate the model upon which those predictions were based.
We will see later that this 1D model works remarkably well.

In our model for longitudinal modes in a 1D medium 
 the desired fields are given by
~
\begin{align}
  \textbf{\textit{E}} &= E(t, z) \underline{\textbf{e}}_{z},~
  \textbf{\textit{P}} = P(t, z) \underline{\textbf{e}}_{z},~
  \textbf{\textit{B}} = \textbf{\textit{H}}
                      = \textbf{\textit{0}} 
\\
 \text{and}\quad
  \textbf{\textit{D}}&= \Permittivity\textbf{\textit{E}}
                     = \Permittivity_{0}\textbf{\textit{E}}+\textbf{\textit{P}}
                     = \textbf{\textit{0}} 
,
\end{align}
 where $\underline{\textbf{e}}_{z}$ is the unit vector along the $z$ axis, 
 our chosen longitudinal orientation.
Maxwell's equation is automatically satisfied
  and the requirement that $\textbf{\textit{D}}=\textbf{\textit{0}}$
  implies that $\Permittivity=0$,
 an ENZ material \cite{Ramaccia-SBT-2013tap}.
Our choice of wire media for our studies was motivated by the existence
 in the literature of a homogenized 1D model for the uniform-radius case,
 which exhibits spatial dispersion.
Such a medium would allow us
 to make the simplifications detailed above,
 as well as possess the required ENZ nature.
However, 
 in order to find a way of controlling the mode profiles we need an
 \emph{inhomogeneous} 1D model: 
 \UPDATED{it needs to be both inhomogeneous
 (i.e. have a position $z$ dependence), 
 as well as 
 spatially dispersive 
 (i.e. have a wavevector $k$ dependence), }
 which is challenging as $k$ and $z$ are
 Fourier conjugate variables
 \cite{Gratus-M-2015jo,Gratus-T-2011jmp,Gratus-KMT-2016njp-stdisp}.
We start with
 the undamped hydrodynamic Lorentz model 
 which is spatially dispersive and homogenous.
The constitutive relation \cite{Belov-MMNSST-2003prb,Song-YSH-2013oe,Ciraci-PS-2013cphc} 
 for this model 
 in the frequency-wavevector representation is 
 usually written as
~
\begin{align}
  \hat{P}(\omega, k)
&=
  -\frac{4 \pi^2 \PlasmaFq^{2} ~ \Permittivity_{0}}
        {\omega^{2} - \omega_0^2 - \Bpolarv^{2}{\cspeed}^{2}k^{2}}\hat{E}(\omega, k) 
.
\label{eqn-Model}
\end{align}
However, 
 in what follows we will use
 ordinary (not angular) frequencies $\wavefq=\omega/2\pi$, 
 and likewise for wavevector $k$ we will use
 the alternative (inverse wavelength) 
 $\waveka = k / 2\pi$.
It follows that in order for $\textbf{P}=-\Permittivity_{0}\textbf{E}$ (i.e. $\textbf{D}=0$) then:
\begin{align}
  \wavefq^2 - \wavefq_0^2 - \Bpolarv^{2}{\cspeed}^{2} \waveka^{2} 
&=
  \PlasmaFq^{2}
\label{eqn-Plasma}
\end{align}
This is a plasma like dispersion relation
 where ${\cspeed}\Bpolarv$ is the polariton velocity,
 $\wavefq_0$ is the polariton resonance frequency, 
 and $\PlasmaFq$ is the plasma frequency parameter. 
If we rearrange \eqref{eqn-Model}
 and Fourier transform {back} into the
 frequency-{physical} space {representation} then we have  
~
\begin{align}
  {\cspeed^{2} \Bpolarv^{2}}
  \frac{\partial^{2}P}
       {\partial z^{2}}
 +
  \left(
    \wavefq^2 - \Cutoff
  \right)
  P
=
  0,
\end{align}
where the cut-off $\Cutofff$
 has units of frequency,
 and is
~
\begin{align}
  \Cutoff
&=
  \wavefq_0^{2}
 + 
  \PlasmaFq^{2}
.
\label{eqn-PermFourier}
\end{align}

Since \eqref{eqn-PermFourier} is written in physical space representation,
 we can now allow for an inhomogeneous permittivity, 
 following \cite{Gratus-M-2015jo}.
This is achieved by allowing for a position dependent
 modulation of the the cut-off,
 which we denote by replacing $\Cutoff$ 
 by $\Cutoffz$.
\UPDATED{Although we might also let the 
 relative polariton velocity $\Bpolarv$ vary, 
 we do not, 
 since as we will show later, 
 for our chosen parameters it stays nearly constant.}
This means that 
 \eqref{eqn-PermFourier} becomes 
\begin{align}
  {\cspeed^{2} \Bpolarv^{2}}
  \frac{\partial^2 P}{\partial {z}^2}
 +
  \left[  \wavefq^2 - \Cutoffz   \right]
  P
&=
  0 
.
\label{eqn-inhomf}
\end{align}
It is not necessary to consider whether the dependence on $z$
 in $\Cutoffz$ 
 is a result of $\wavefq_0$ or $\PlasmaFq$ having a $z$ dependence,
 or of both. 
We can now see that \eqref{eqn-inhomf} can be made
 equivalent to Mathieu's equation \eqref{eqn-Mathieu},
 simply by
 rescaling the spatial coordinate with $\sigma=2\pi z / \Length$,
 and setting the cut-off modulation $\Cutoffz$ to be
\begin{align}
f^2 - \Cutoffz
&=
  \frac{ {\pfa} \cspeed^{2} \Bpolarv^{2}}
       {\Length^2}
  \left[  a - 2 q \cos\left( 4 \pi z / \Length \right)  \right]
.
\label{eqn-Lambda}
\end{align}
~
By doing this we have reconciled the equations of our wire media
 with Mathieu's equation, 
 a process which required making either the
 plasma frequency $\PlasmaFq$ or the resonance frequency $\wavefq_0$
 dependent on $z$.
This can be achieved by realizing that these frequencies
 depend on the radius $\Radius$ of the wires in the wire media.
Our goal therefore is to find the appropriate variation
 in the wire radius $\Radius$, 
 which we will denote $\Radiusz$, 
 to implement in our 3D simulations. 
In section \ref{S-3Dnumerics}
  below we use set of full 3D simulations with uniform wires to 
 calculate the dependence of dispersion relation \eqref{eqn-Plasma}
 on the radius.
Thus 
 for some specified wire radius $\Radius$,
 the cut-off value is obtained by numerical simulation; 
 where of course this simulated cut-off,
 denoted $\CutofffSimr$,
 is in turn dependent on how the frequencies $\PlasmaFq$ and
 $\wavefq_0$ vary with the wire radius $\Radius$.
Hence \eqref{eqn-Plasma} can be rewritten as
\begin{align}
  \Bpolarv^2 
  {\cspeed}^2
   \waveka^2 
&=
  \wavefq^2
 -
  \CutoffSimr
.
\label{eqn-b2c2k2}
\end{align}
In order to find the the radius variation $\Radiusz$ we simply solve
\begin{align}
\CutoffSim(\Radiusz) = \Cutoffz
,
\label{eqn-RadiusVariationR}
\end{align}
where $\CutoffSimr$ is given by \eqref{eqn-b2c2k2}
 and $\Cutoffz$ is given by \eqref{eqn-Lambda}\footnote{\UPDATED{We will
  see in the next section 
  that $\CutofffSimr$ follows
  a simple and invertible fitting formula \eqref{eqn-Approx},
  meaning that we can solve for \eqref{eqn-RadiusVariationR} analytically.}}

\begin{figure}
\centering
\resizebox{0.99\columnwidth}{!}{
\input{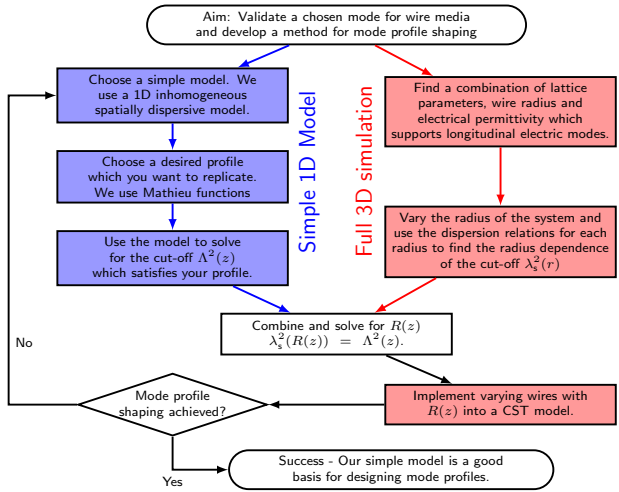}
}
\caption{The steps used for generating and then validating
 our field profiling model.}
\label{fig-Flowchart}
\end{figure}

We now need to combine these results with a 3D simulation of our structure. 
The flowchart shown in fig. \ref{fig-Flowchart}
 shows our method for designing 3D structures
 on the basis of our 1D predictions. 
It requires that we find longitudinal plasma-like modes
 in uniform-radius 3D simulations which persist for a range of wire radii. 
This gives us the cut-off values
 as a function of the wire radius $\CutoffSim(\Radius)$,
 and we can match these up with the 1D model's 
 required modulation in the cut-off.
The result will be that we can find the necessary wire radius
 as a function of position for a full 3D mode-profiling simulation. 
The next section will describe the uniform-radius 3D simulations
 that give us $\CutoffSimr$.

\section{3D numerical calculations of mode dispersion: \textrm{$\CutoffSimr$}}\label{S-3Dnumerics}

\begin{figure}[h]
\centering
\includegraphics[width=0.65\linewidth]{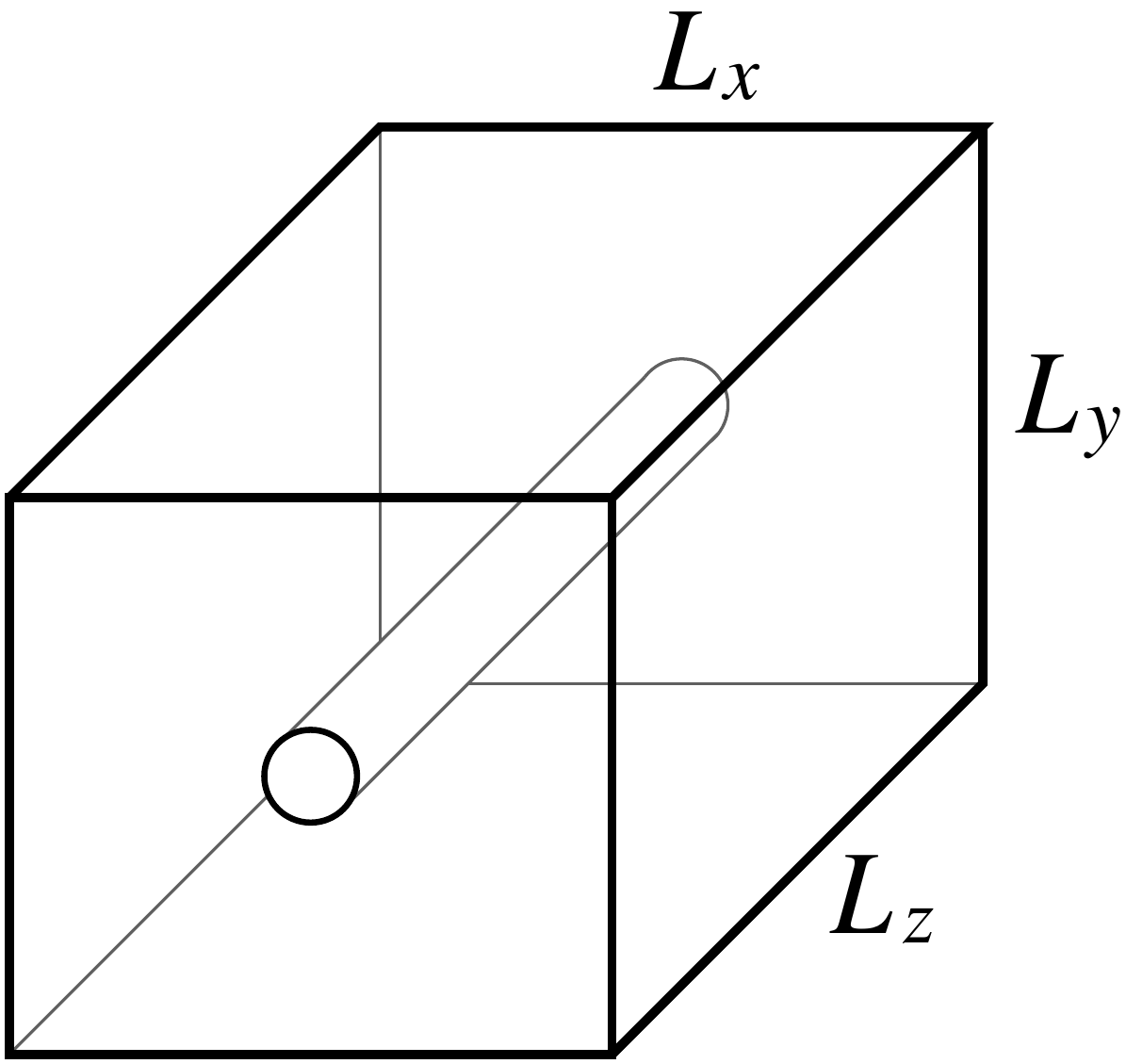}
\caption{A 3D box with a uniform-radius wire 
 and the unit-cell aspect ratio we use.
The transverse dimensions are close to, 
 but not exactly square, 
 with $\Length_x=15\Units{mm}$, 
 and $\Length_y=13.06\Units{mm}$.
\UPDATED{The longitudinal depth 
 (i.e. the wire/cell length)
 used does not affect $\CutoffSimr$, 
 but does allow us to
 specify a a convenient $\waveka$ value
 where the 
 band folding occurs.}
}
\label{fig-3Dunitcell}
\end{figure}

Here we find longitudinal electric modes
 with a plasma-like dispersion relation
 in uniform-radius 3D simulations which persist for a range of wire radii. 
These CST simulations were based on 
 a single unit cell as shown on fig. \ref{fig-3Dunitcell},
 containing a dielectric rod of radius $\Radius$, 
 and with periodic boundary conditions in all directions,
 thus representing an infinite wire media.
The calculations in this section all incorporate cells of this type, 
 but with differing wire radii $\Radius$,
 so that we can find $\Cutoffr$
 for the range of 3D structures we are interested in.

Our standard (reference) simulation was for
 an infinite dielectric wire medium,
 where the wires had a permittivity
 of $\Permittivity=1600\Permittivity_{0}$ 
 (e.g. ceramics \cite{Li-KWT-2008je})
 and a radius of $0.2\Units{mm}$, 
 with lattice constants of $15\Units{mm}$ and $13.06\Units{mm}$
 in the $x$ and $y$ (transverse) directions respectively. 
This use of a non-square unit cell cross-section ensures there is no 
 ambiguity due to degeneracy between $x$ and $y$ orientations 
 in the simulation output.
Note that in our 3D description, 
 $\waveka$ is retained (solely) as
 the Fourier conjugate variable to the longitudinal spatial axis $z$.
\UPDATED{Transverse wavevectors do play a role in that they 
 help determine the mode shapes, 
 and in particular whether or not any given mode
 has our desired longitudinal property.
As such 
 they are dependent on $1/\Length_x$ and $1/\Length_y$, 
 but their specific values play no role in our calculations here.}

\begin{figure}
\centering
\includegraphics[width=0.80\columnwidth]{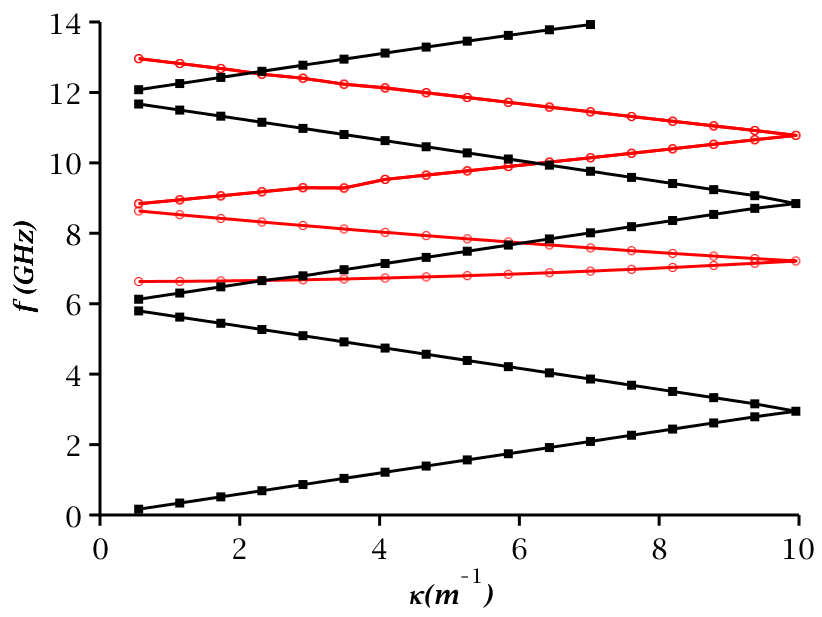}
\caption{\UPDATES{The dispersion properties
 of a uniform-radius wire
 with $\Radius=0.2\Units{mm}$,
 and length $\Length_z=50.25\Units{mm}$;
 $\Permittivity=1600$.
Multiple bands are seen.
The black squares show results for the transverse modes, 
 where the field is primarily oriented perpendicular to the wires; 
 in such modes the wires are nearly invisible to the electroagnetic field, 
 as is shown by their vacuum-like dispersion.
The red circles show results for the longitudinal modes, 
 where the field is primarily oriented parallel to the wires.
The numerical data is joined together by lines as a guide for the eye.}}
\label{fig-Bandstructure0}
\end{figure}


\begin{figure}
\centering
\includegraphics[width=0.99\columnwidth]{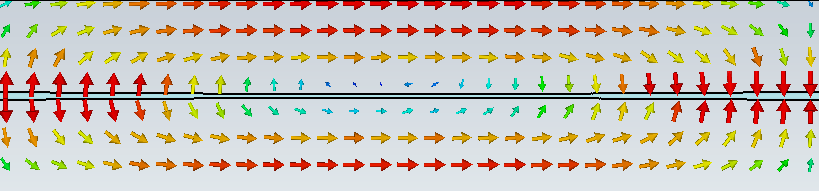}\\
\includegraphics[angle=90,height=0.34\columnwidth]{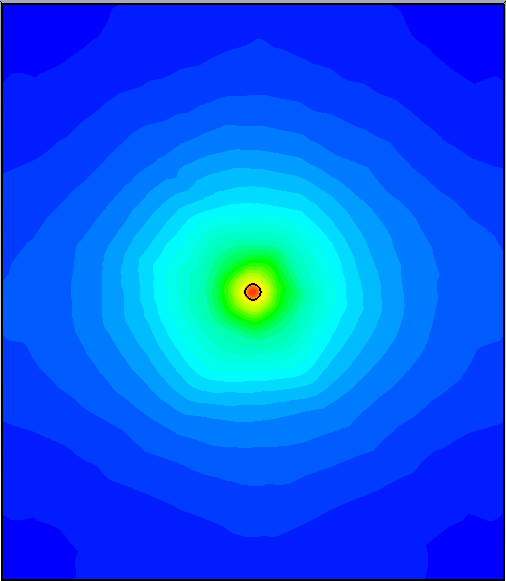}
\hfill
\includegraphics[angle=90,height=0.34\columnwidth]{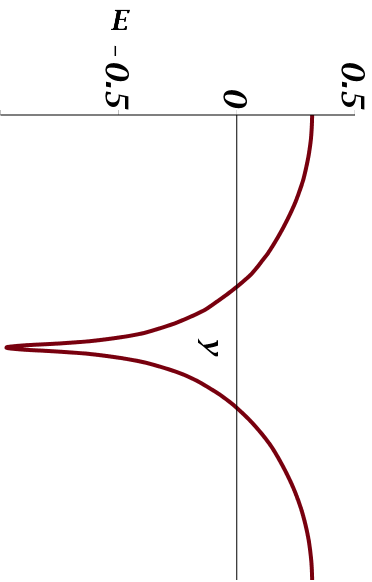}
\caption{Images showing
 \UPDATES{(top) 
 the field strength and orientation 
 in a cross-section of one half-length of the unit cell
 for 
 a longitudinal mode;
 (bottom left) 
 the \emph{transverse} variation of the longitudinal field component, 
 (bottom right) 
 the same \emph{transverse} variation of the longitudinal field component, 
 along a line crossing the unit cell.}
}
\label{fig-LongPlots}
\end{figure}


These simulations, 
 as well as demonstrating the expected predominance of transverse modes, 
 confirmed the existence of longitudinal electric modes
 in wire media structures. 
The longitudinal modes found for the parameters above
 have a high band index\footnote{Determined by 
  counting the bands found by CST
  from lowest frequency to highest.
  However, 
   note that band index determination in numerical results 
   is often complicated 
   or made ambiguous due to band crossings
   and the choice of unit cell sizing.}
 and a cut-off frequency of about 
 ${\CutofffSim} = \UPDATES{6}\Units{GHz}$,
 as can be seen on fig. \ref{fig-Bandstructure0}.
However,
 note that the index of the longitudinal mode
 differs for different simulation parameters, 
 since the changing geometry
 has an effect on the details of the bandstructure.

These longitudinal modes
 have the significant features that the field between the wires
 was both strongly longitudinal and anti-parallel to the field in the wires,
 as shown in fig. \ref{fig-LongPlots}.
This is ideal for applications where the resulting field
 is intended to accelerate or decelerate particles along their path
 in a free space region.

\begin{figure}
\centering
\includegraphics[width=0.9\linewidth]{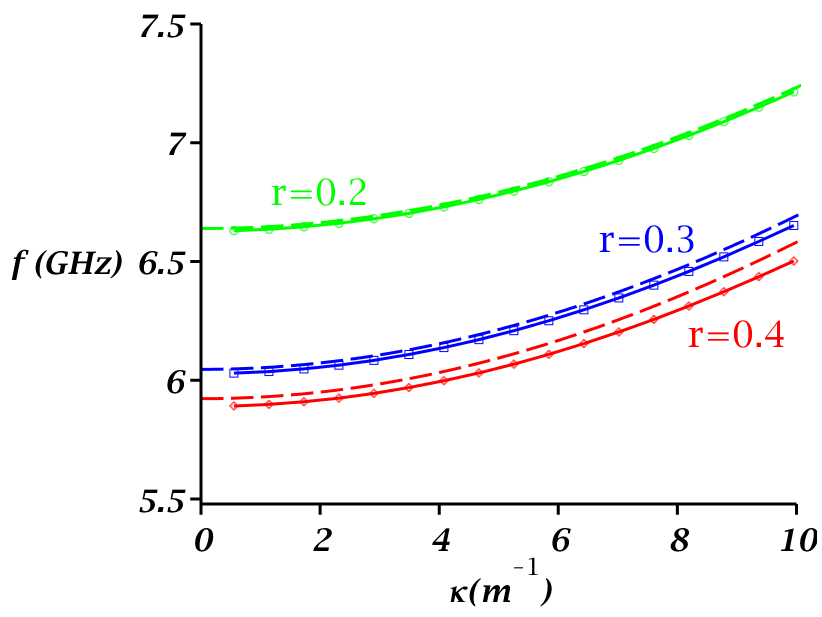}
\caption{Dispersion plot
 for the chosen longitudinal mode present
 in our standard uniform-radius dielectric wire medium
 with $\Permittivity=1600$.
We see the characteristic parabolic shape 
 typical of a plasma-like response.
The $y$-intercept of this figure indicates the cut-off ${\CutofffSim}$.
\UPDATES{We found that that for wires of radius 
 $\Radius = ~0.20\Units{mm}$ (black, circles) that
  the cut-off is $\CutofffSim \simeq 6.64\Units{GHz}$; 
 that for 
 $\Radius = ~0.30\Units{mm}$ (blue, squares) that
  the cut-off is $\CutofffSim \simeq 6.05\Units{GHz}$; 
 and
 $\Radius = ~0.40\Units{mm}$ (red, diamonds) that 
  the cut-off is $\CutofffSim \simeq 5.92\Units{GHz}$.}
\UPDATES{
The solid lines indicate the parabolic fits to the dispersion data; 
 whereas the dashed lines show the predicted dispersion 
 made on the basis of the later phenomenological fit
 of \eqref{eqn-Approx}.}}
\label{fig-DispersionPlot} 
\end{figure}

The dispersion properties of these longitudinal modes
 are shown in fig. \ref{fig-DispersionPlot},
 where we see that they obey the plasma-like dispersion relation
 predicted by theoretical analysis. 
This was confirmed by the plotting of $\wavefq^{2}$ vs. $\waveka^{2}$ graphs
 which show a straight line,
 from which the cut-off $\CutoffSim$,
 and polariton velocity $\Bpolarv$,
 can be straightforwardly obtained, 
 although 
 $\wavefq$ vs. $\waveka$ graphs
 are often plotted instead.
In our simulations, 
 we found that the dominant effect of radius variation
 is on $\CutoffSim$, 
 with $\Bpolarv$ being  
 approximately constant for all simulations at $\Bpolarv=0.96$.

\label{fig-Modes}

\begin{figure}[h]
\centering
\includegraphics[width=0.9\linewidth]{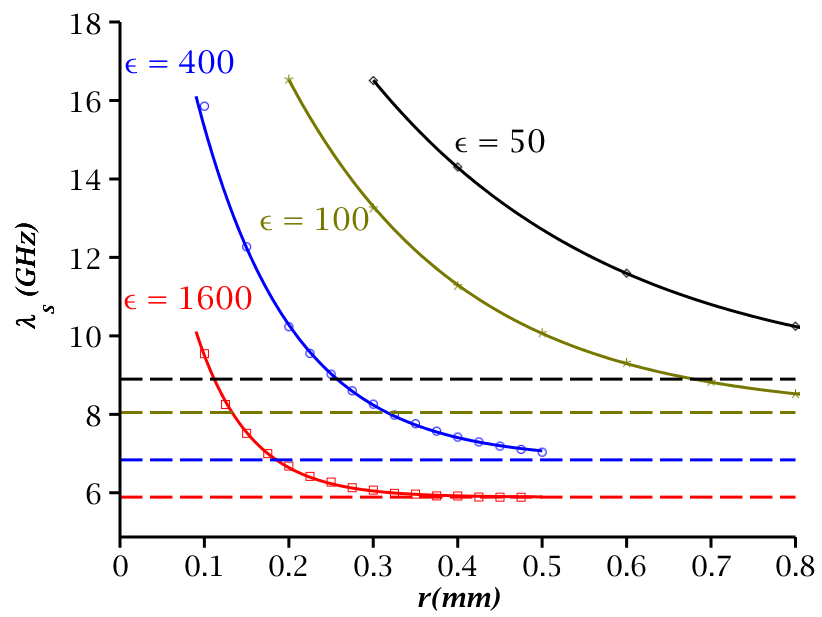}
\caption{The dependence
 of $\CutofffSim$ 
 on the radius $\Radius$ of a uniform wire.
Note that the vertical axis does not extend down to $\CutofffSim=0$. 
\UPDATED{Data is shown for four cases, 
 with different permittivity values
 (i.e. $\Permittivity \in \{ 50, 100, 400, 1600\}$); 
 where data points for for smaller radii than those shown
 significantly affected the quality of the otherwise remarkably good fit,
 so we excluded such points from our analysis.
The dashed lines indicate the large radius (asymptotic)
 value for $\CutofffSim$, 
 which gives the value of $C$ in \eqref{eqn-Approx}.}
}
\label{fig-PlasmaComp} 
\end{figure}

We now use our simulations to find the dispersion relations
 for a useful range of wire radii, 
 and thus find (by fitting) the corresponding cut-offs. 
On fig. \ref{fig-PlasmaComp} we can see how these change with wire radius,
 and by fitting this data,
 it is the possible to find a relation
 for the radius dependence of the cut-off:
~
\begin{equation}
  \CutofffSimr
\approx
  C
 +
  A 
  \exp\left(  -\frac{\Radius}{\rho}  \right)
.
\label{eqn-Approx}
\end{equation}
Here $C$, $A$ and $\rho$ are fitting constants 
 that depend on the wire permittivity, 
 unit cell, 
 and mode index.
Although this relation is only approximate,
 the agreement should be sufficient
 so long as we stay within the range of data taken for the radius,
 between $0.1\Units{mm}$ and $0.5\Units{mm}$. 
As can be seen on fig. \ref{fig-PlasmaComp},
 these parameter fits 
 are remarkably accurate, 
 with \emph{exceptionally} small deviations from the model behaviour.
\UPDATES{For example, 
 we can see on fig. \ref{fig-DispersionPlot} 
 the small discrepancy 
 between the dispersion relation predicted by this fit, 
 and that from the numerical data on which it was originally based.
This difference is negligible on the scale of the results 
 on fig. \ref{fig-PlasmaComp}, 
 but increases for larger radii.}

We repeated this process for thicker rods with lower permittivities, 
 an important consideration when we wish to move towards
 real-world fabrication possibilities.
We verified that mode profiling was still viable
 at lower relative permittivities, 
 i.e. at both $\Permittivity = 50$,
 $\Permittivity = 100$,
 and $\Permittivity = 400$. 
As can be seen in fig. \ref{fig-PlasmaComp},
 for a given frequency,
 the radius required for longitudinal modes to appear
 becomes larger for a lower relative permittivity.
This means that the mode profiling structures for $\Permittivity = 50$ 
 involve larger wire radii in the order of millimetres.
Such lower index materials are not only easier to find, 
 but may be very useful in practical applications where
 such thin rods may be either difficult to fabricate
 or too fragile.
For the results shown on fig. \ref{fig-PlasmaComp}, 
 the fitting parameters for \eqref{eqn-Approx}
 are shown on table \ref{table-fitting}.

%

\begin{table}[h]
\centering
 \begin{tabular}{|c|c|c|c|c|c|}
 \hline
 $\Permittivity$  &  $A$  &  $C$  &  $\rho$  &  $\chi^{2*}$  &  R$^2$ \\
 \hline
 \hline
    1600  &  17.3  & 5.89  &  0.0637  &  $1.08 \times 10^{-3}$ & $0.9991$ \\
    ~400  &  18.5  & 6.77  &  0.1193  &  $1.01 \times 10^{-3}$ & $0.9999$ \\
    ~100  &  22.1  & 8.04  &  0.2088  &  $0.84 \times 10^{-3}$ & $0.9999$ \\
    ~~50  &  21.5  & 8.87  &  0.2901  &  $0.49 \times 10^{-3}$ & $0.9999$ \\
 \hline
 \end{tabular}
\caption{
The fitting parameters $A$, $C$, and $\rho$
 for each choice of wire permittivity $\Permittivity$;
 where the values of $\chi^{2*}$ and R$^2$
 indicate the exceptionally good fit to the model behaviour
 proposed in \eqref{eqn-Approx}.
}
\label{table-fitting}
\end{table}

%
%
%


\section{The wire shape needed to match a Mathieu equation: $\Radiusz$}
\label{Profiling}

We now need to 
 describe how to match the properties of our wire media, 
 and the fields they can support, 
 to the mathematical form of Mathieu's equation.
This comparison will tell us how to turn a mathematical Mathieu profile
 into the wire radius variation that will support it.
\UPDATED{Inserting the definition \eqref{eqn-RadiusVariationR}
 into \eqref{eqn-Lambda}
 allows us to see that we simply require}
~
\begin{align}
  \CutoffSim\left(\Radiusz\right)
&=
  \wavefq^{2}
 -
  \frac{4\pi^{2} \Bpolarv^{2} \cspeed^{2}}{\Length^{2}}
  \left[
    a
   -
    2 q \cos\left(\frac{4\pi z}{\Length}\right)
  \right]
.
\label{eqn-Master}
\end{align}
\UPDATED{The factors $\Bpolarv$, $a$, $q$
 and the parameters which define $\CutoffSim(\Radius)$
 are determined either by the simulations or by the choice of profile shape. 
We see that to determine $\Radiusz$ from \eqref{eqn-Master}
 we need only specify the frequency $\wavefq$
 and the repetition length $\Length$; 
 but although 
 we have a lot of freedom in our choice of $\Length$ and $\wavefq$,
 there are some constraints. 
Notably, 
 by taking the maximum and minimum values of the cosine function
 we obtain the concomittant 
 maximum and minimum values of the cut-off, 
 being
~
\begin{align}
  \Cutoff_{\pm}
&=
  \wavefq^{2}
 +
  \frac{4\pi^{2}\Bpolarv^{2}\cspeed^{2}}
       {\Length^{2}}
  \left( a \pm 2q \right)
.
\label{eqn-Two}
\end{align}
From figure \ref{fig-PlasmaComp} we see that $\CutoffSim(\Radius)$ is bounded
 so we require $\Cutoff_{-} > C$, 
 where $C$ is given by \eqref{eqn-Approx}. 
Further, 
 the requirements resulting from actually constructing 
 the wire medium device may also 
 demand a value for the minimum radius
 which will then in turn constrain $\Cutoff_{+}$.}

\UPDATED{From \eqref{eqn-Two} we see
 there is an inverse relationship between
 the range of $\Cutoff$ and the length.
\begin{align}
  \Length^2
&=
  \frac{16 \pi^{2} \Bpolarv^{2} \cspeed^{2}q}
       {\Cutoff_{+} - \Cutoff_{-}}
\end{align}
Thus the same overall mode shape
 can be designed to have different periodicities, 
 and be made with (e.g.) 
 either a short $\Length$ and a large cut-off modulation,
 or a long $\Length$ and a small modulation.}

\begin{figure}
\includegraphics[width=0.90\columnwidth]{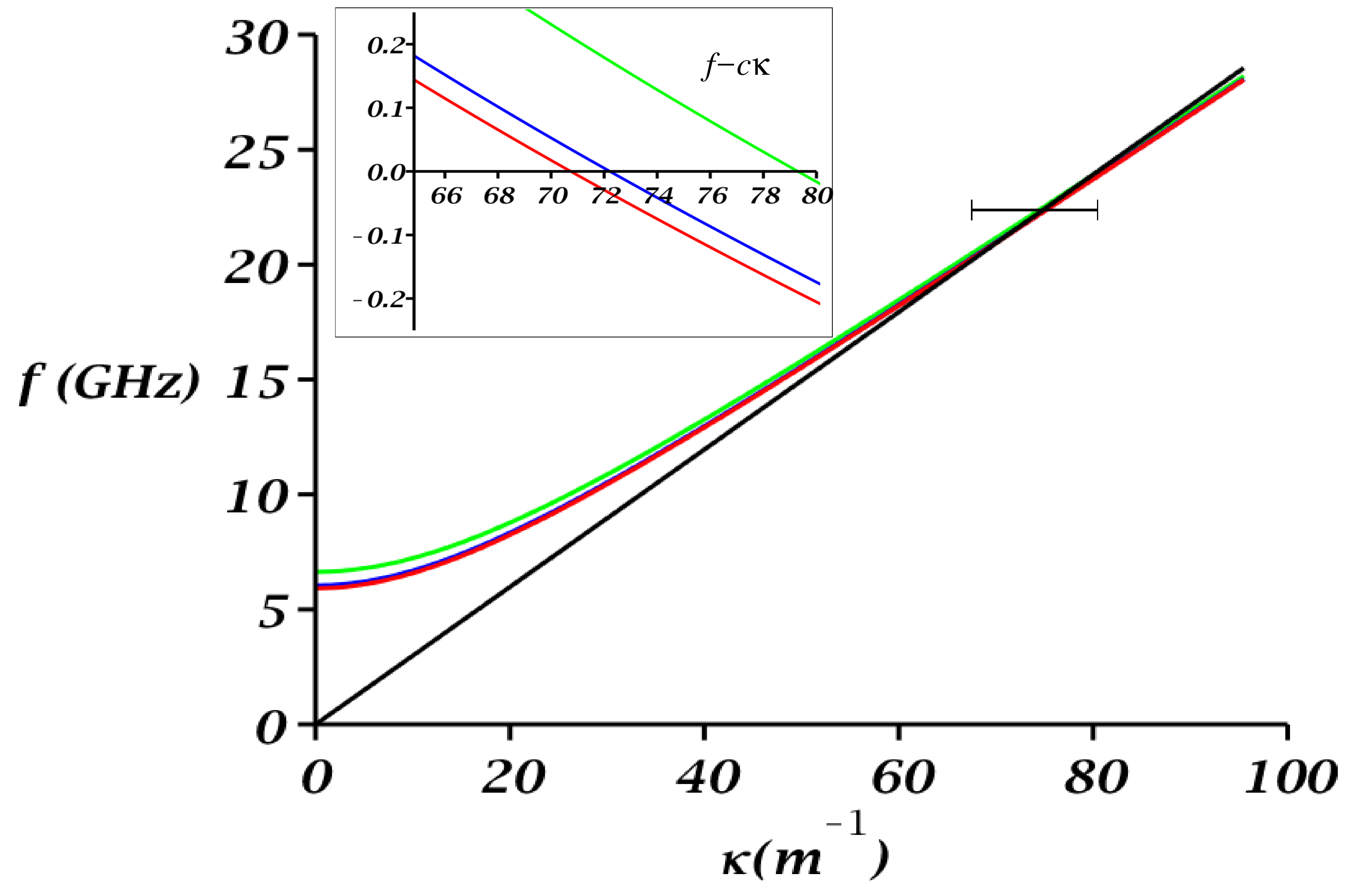}
\caption{Predicted
  dispersion relations from \eqref{eqn-b2c2k2},
 with different radii, 
 and with the vacuum speed of light in black.
This is a replotted version of fig. \ref{fig-DispersionPlot}, 
 but over a wider range of $\waveka$ and $\wavefq$.
Assuming the dispersion relation holds for such high frequency
 then for $\wavefq$ between about $25\Units{GHz}$ and $35\Units{GHz}$
 the phase velocity equals the speed of light. 
However such high $\waveka$ values correspond to very short wavelengths
 where the plasma dispersion relation may break down.
\UPDATES{The inset, 
 covering the range of $\waveka$ indicated by the horizontal bar, 
 plots the difference between the wire media dispersions
 and the vacuum dispersion
 (i.e. $f-\cspeed\waveka$), 
 and shows how the various dispersia cross at high frequencies.}}
\label{fig-PredictedDispersion}
\end{figure}

\UPDATED{For the work presented here 
 we choose a frequency
 near that of the cut-off, 
 which means that our field profiles are in effect standing waves.}
\UPDATES{However, 
 there is the possibility of investigating 
 whether we can construct propagating field profiles, 
 i.e. ones which have a finite phase velocity, 
 by working higher up the dispersion curves.
Since $\Bpolarv < 0$,
 then for small $\waveka$,
 the waves are superluminal;
 whereas for large $\waveka$ the waves are instead subluminal. 
In figure \ref{fig-PredictedDispersion} we see that 
 setting $\wavefq=\UPDATES{20}\Units{GHz}$
 then one obtains a phase velocity of about $\cspeed$. 
Although we would then expect the profiles to also become time dependent, 
 such an achievement would bring us closer to our goal
 of matching the wave profile speed to the particle speed\footnote{However, 
  due to the extra complications and more demanding computations involved,
  this is beyond the the parameter ranges we have examined to date,
  and such high modes may not have the desired hyperbolic response.}, 
 enhancing both their interaction
 and the utility of that interaction. 
Note however that since our modes are not sinusoidal,
 and are therefore consist of several spatial harmonics,
 one can no longer easily assign a single phase velocity.}


\begin{figure}
\centering
\includegraphics[width=0.80\columnwidth]{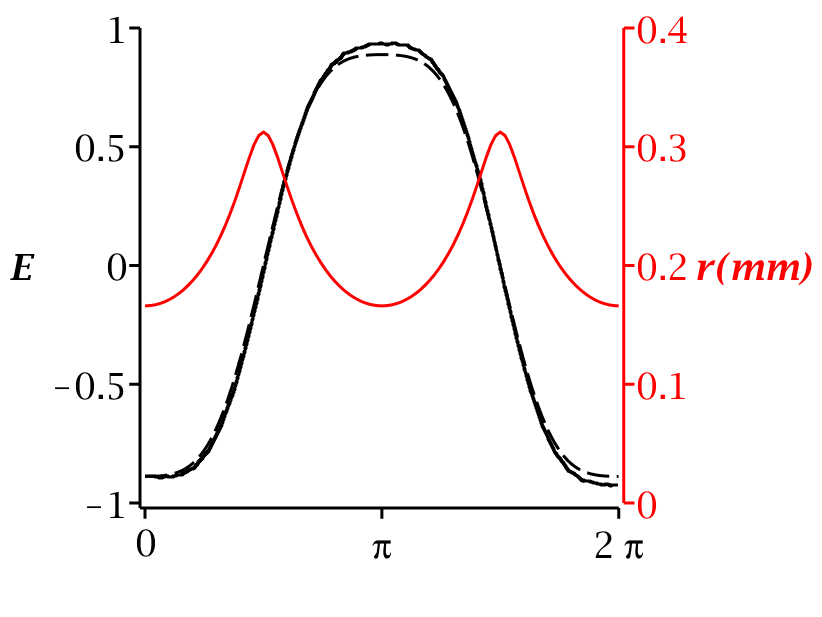}
\caption{\UPDATES{3D frequency domain CST simulation results for
 a single unit cell, 
 showing 
 the flat field profile (solid line) 
 and the aimed-for Mathieu solution (dashed line), 
 as well as 
 (red line, right-hand scale)
 the double-peaked radius variation used to generate the field profile.
These were based
 on a cell length of $\Length_{z,\textup{flat}}= ~132 \Units{mm}$
 a frequency of $f_{\textup{flat}}= ~7.2 \Units{GHz}$,
 and a cut-off variation with $\Cutofff_\pm =  ~6.0, ~7.2 \Units{GHz}$.
}}
\label{fig-Results-Flat}
\end{figure}

\begin{figure}
\centering
\includegraphics[width=0.80\columnwidth]{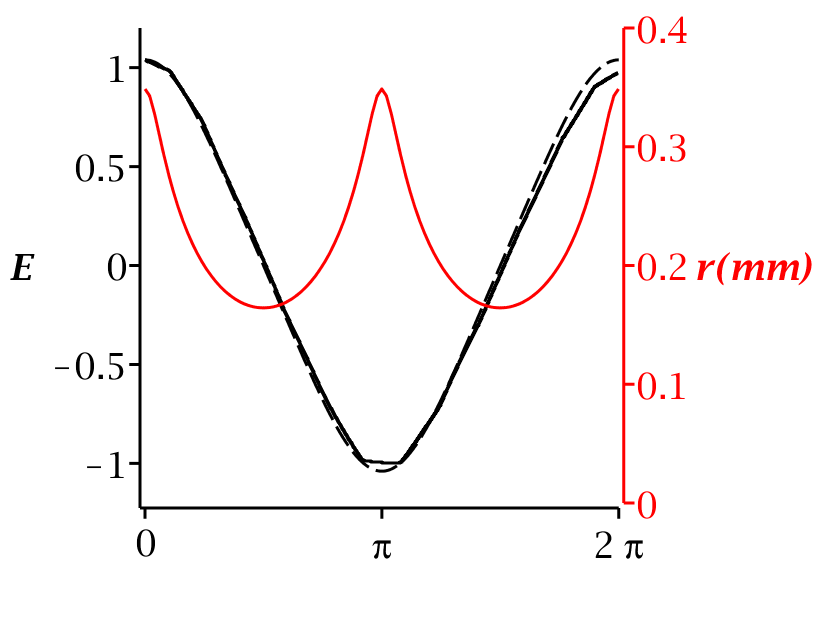} 
\caption{\UPDATES{3D frequency domain CST simulation results for
 a single unit cell, 
 showing 
 the triangular field profile (solid line) 
 and the aimed-for Mathieu solution (dashed line), 
 as well as
 (red line, right-hand scale)
 the double-peaked radius variation used to generate the field profile.
These were based
 on a cell length of $\Length_{z,\textup{tri}}= ~82 \Units{mm}$
 a frequency of $f_{\textup{tri}}= ~7.3 \Units{GHz}$,
 and a cut-off variation with $\Cutofff_\pm = ~6.0, ~7.2 \Units{GHz}$.
}}
\label{fig-Results-Tri}
\end{figure}

\begin{figure}
\centering
\includegraphics[width=0.80\columnwidth]{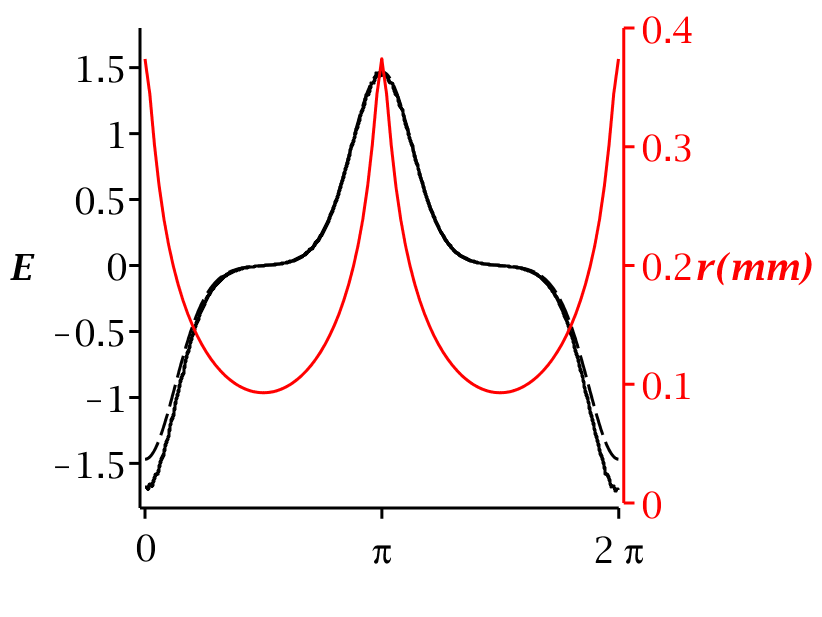} 
\caption{\UPDATES{3D frequency domain CST simulation results for
 a single unit cell, 
 showing 
 the peaked field profile (solid line) 
 and the aimed-for Mathieu solution (dashed line), 
 as well as
 (red line, right-hand scale)
 the double-peaked radius variation used to generate the field profile.
These were based
 on a cell length of $\Length_{z,\textup{peak}}= ~691 \Units{mm}$
 a frequency of $f_{\textup{peak}}= ~7.3 \Units{GHz}$,
 and a cut-off variation with $\Cutofff_\pm = ~6.0, ~9.9 \Units{GHz}$.
Note that this profile shape requires much longer cell lengths
 than the others, 
 as a result of its $a, q$ Mathieu parameters.
}}
\label{fig-Results-Peak}
\end{figure}


Now that 
 we have a method for determining the remaining two free parameters,
 $\wavefq$ and $\Length$,
 it is now possible to solve \eqref{eqn-Master} for $\Radiusz$. 
This has the form
~
\begin{align}
  \Radiusz
&=
 -
  \Gamma_1
  \ln
    \left\{
        \Gamma_2
       +
        \Gamma_3
        \cos( \Gamma_4 z)
    \right\}
,
\label{eqn-rz-natural}
\\
\textrm{where}\qquad
 \Gamma_1 &= \rho
,
\\
 \Gamma_2 &=
    \frac{
        f_0^2 \Length^2 
       -
        a \Bpolarv^2 {\cspeed}^2
       -
        C \Length^2 
    }
    {A \Length^2}
,
\\
 \Gamma_3 &=
        \frac{2 \Bpolarv^2 {\cspeed}^2 q}{A \Length^2}
,
\\
 \Gamma_4 &=4\pi/\Length
.
\label{eqn-rz-Gamma}
\end{align}
To generate the radius variation needed for each
 of our selected mode profiles, 
 and 
 for our standard $\Permittivity=1600$ wire medium,
 the necessary $\Gamma_i$ parameters are given by 
 table \ref{table-radiusz}.
Implementing this $\Radiusz$ function in a dielectric wire medium
 should then satisfy \eqref{eqn-Master},
 allowing for the support of longitudinal modes with the desired mode profiles, based on the chosen Mathieu function.


\begin{table}
 \begin{tabular}{|c|c|c|c|c|}
 \hline
 Shape & $\Gamma_2$  &  $\Gamma_3$  &  $\Gamma_4$ \\
 \& frequency  & & &  \\
 \hline
 \hline
   Flattop    & 0.02474455655 & +0.02164881399 & 95.00527368 \\
     $7.2\Units{GHz}$   & & &  \\
 \hline
   Triangular & 0.02474455654 & -0.02316162974 & 153.2365130 \\
     $7.3\Units{GHz}$   & & &  \\
 \hline
   Peaked     & 0.02474455657 & -0.02417189946 & 28.39429186  \\
     $6.5\Units{GHz}$   & & &  \\
 \hline
 \end{tabular}
\caption{
The parameters specifying the radius variation 
 for the three chosen longitudinal field profiles, 
 for use in \eqref{eqn-rz-Gamma}.
For each of the three profiles, 
 $\Gamma_1 = 0.5406647836\Units{mm}$ for all three shapes.
}
\label{table-radiusz}
\end{table}






\section{Results and Discussion}

We followed the methodology outlined above to create
 varying wire unit cells for our three different wave profiles.
The results of these simulations are shown
 on figs. \ref{fig-Results-Flat}, \ref{fig-Results-Tri},
 \ref{fig-Results-Peak},
 and they confirm that our method accurately reproduces
 the desired longitudinal field profiles 
 given the correctly specified wire media structures.
As $\Radiusz$ will be a periodic function
 it is fairly simple to implement this profile in a unit cell
 with periodic boundary conditions in CST, 
 although it should be noted
 that a feature of Mathieu's equation means that
 the period (wavelength) of the Mathieu mode will be $\Length$,
 which is double the period of the wire variation. 
Though an exact replication of the radius variation is not possible in CST
 it is possible to make an approximation of the variation
 with a large number of connected conical frustums.

%

We have seen above, 
 notably in fig. \ref{fig-PlasmaComp},
 that despite our initial preference for 
 very thin high permittivity wires, 
 our approach still has currency 
 even for thicker wires with lower permittivity.
We are currently checking how far this process can be pushed, 
 i.e. to what base wire thickness and what low permittivity.

\begin{figure}
\centering
\includegraphics[width=0.80\columnwidth]{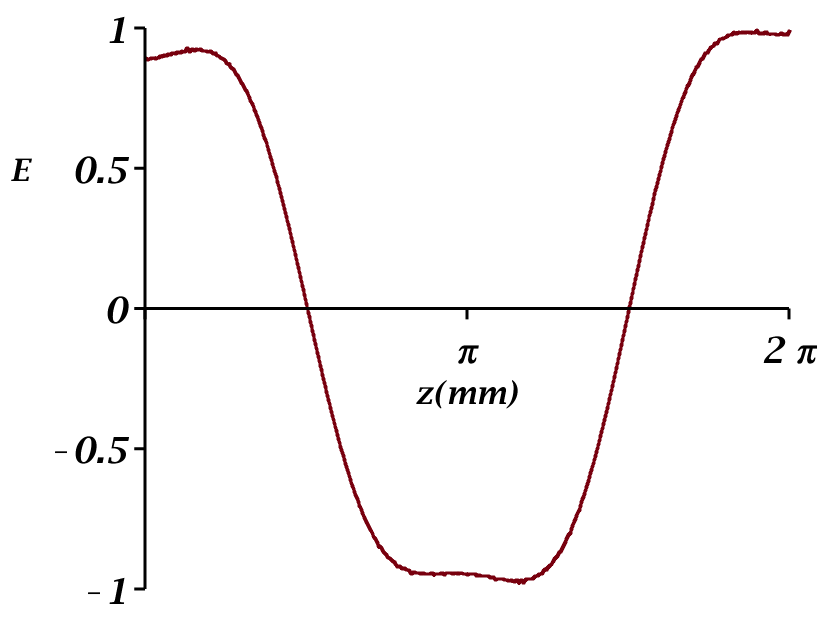}
\caption{\UPDATES{The flat-top field profile
 obtained using a simulation with a coated dielectric wire
 with $L=317.4\Units{mm}$ and permittivity $\Permittivity=50$, 
 whose profile was based on wire data for an \emph{unclad} wire, 
 and varied between $0.69\Units{mm}$ and $0.71\Units{mm}$.
The cladding covers over the dielectric wire
 out to an outer radius of $3\Units{mm}$,
 and has a relative permittivity of $\Permittivity = 2$.
}}
\label{fig-coatedwires}
\end{figure}

\UPDATES{Note that 
 even materials that would be entirely impractical 
 for wires if used on their own --
 such as BaTiO3-based ceramics \cite{Li-KWT-2008je}, 
 which can have $\Permittivity \sim 1000+$ in the \Units{GHz} --
 might still be used if encapsulated in an outer sheath
 of more tractable material.
Simulations have shown 
 that adding a low-index sheath 
 over the high-index core with varying radius 
 does not change the outcome of our 
 (uniform wire) simulations significantly.
We can see in the results on fig. \ref{fig-coatedwires}
 that the dispersion properties of the wire media
 can be sufficiently unaffected by the sheath, 
 so that as a consequence the profile shaping
 will also still persist.}

\UPDATED{We are currently working on extending our numerical results
 into the time domain, 
 either with external driving or internal excitation.
However, 
 obtaining such results can be problematic,
 due to the greatly increased computational requirements.
Such simulations also require a specification
 for how to drive the mode profiles, 
 as indeed is also needed for eventual experiments.
Preliminary indications are that driving our mode profiling waveguides
 externally might be difficult due to poor input coupling.
One resolution of this is to place a dipole antenna \emph{internally},
 and so drive from inside the medium,
 as certainly worked well for our slotted medium mode profiling investigations
 \cite{Gratus-KLB-2017apa-malaga,Gratus-PLB-2017jpc}.
Indeed, 
 it is very convenient that 
 our goal of achieving free-space mode profiling
 makes such a step so simple.}

\UPDATED{Beyond these theoretical extensions, 
 two more steps need to be taken to make our concept application ready.
First, 
 we need to consider the actual structure of a device, 
 which will most likely consist of a finite wire array 
 housed in a metallic box.
Since the field patterns present here have strong components --
 albeit of opposite sign --
 both in the wire and in free space, 
 the best positioning for the box walls with respect to the wires
 needs to be investigated.
Second, 
 we have currently focussed on standing wave field profiling, 
 rather than the travelling wave profiles needed for 
 our aim of finding accelerator applications.
However, 
 now we have established that field profiling is possible 
 with a high degree of accuracy, 
 we are testing configurations aimed at building the wire media 
 using a mode basis that is not centred around 
 near cut-off frequencies and hence low $\waveka$ values
 in the dispersion curves, 
 but at a median $\waveka$ values which have a faster group velocity.}

\UPDATES{Our next step 
 will be to add time domain shaping to our toolbox, 
 either in concert with the spatial profiling shown here, 
 or by itself.
This would use multiple frequency components, 
 and greatly enhance the possibilities for the bunch control
 of charged particles in accelerator applications.}

%

\section{Conclusion}

In this paper we have extended the theoretical analysis
 for metal wire media to dielectric wire media; 
 with both reducing to a 1D homogeneous model incorporating spatial dispersion.
We have shown that these dielectric media support modes
 with longitudinal electric fields 
 (i.e. field parallel to the wires)
 with a plasma-like dispersion relation; 
 and how the plasma frequency $\PlasmaFq$ and mode cut-off $\Cutoff$
 depends on the chosen radius for the wires.

This then formed a basis for our method 
 incorporating a varying wire radius, 
 and allowed us to show
 how the resulting dispersion relation for the modes
 could be manipulated into corresponding to Mathieu's equation. 
We could then create mode profiles 
 which replicated solutions to Mathieu's equation; 
 thus providing an effective approach for implementing 
 specific Mathieu functions (solutions) as physical electric field profiles.
As part of the process we required a fit for the 
 dependence of the cut-off on the wire radius, 
 and we found that a simple exponential model gave remarkable agreement, 
 greatly simplifying the prediction of the wire radial variations needed.
\UPDATES{The exceptional accuracy of this fitting function suggests that 
 a theoretical investigation should yield a useful and accurate
 radius-dependent simplified model of spatial dispersion, 
 making it possible to dispense with our 
 3D uniform-radius unit-cell CST simulations.}

We applied our method to a selection of  chosen mode profiles, 
 namely a flat-top, peaked, and triangular waveforms.
The results of the frequency-domain simulations for these examples
 showed very good agreement, 
\UPDATES{not only for straightforward single unit-cell calculations, 
 but also for wires stabilised in an external cladding.
We also discussed routes for future work, 
 including time domain simulations, 
 investigation of options for exciting/driving the mode profiles, 
 device-like simulations involving multiple wires in a waveguide, 
 and the possibilities for group velocity engineering of the modes.}

\section*{Funding}
 STFC (the Cockcroft Institute ST/G008248/1 and ST/P002056/1)\\
 EPSRC (the Alpha-X project EP/J018171/1 and EP/N028694/1)

\section*{Acknowledgments}
Paul Kinsler would like acknowledge the hospitality of 
 Imperial College London.


\end{document}